\documentclass[fleqn,usenatbib]{mnras}

\usepackage[T1]{fontenc}

\DeclareRobustCommand{\VAN}[3]{#2}
\let\VANthebibliography\thebibliography
\def\thebibliography{\DeclareRobustCommand{\VAN}[3]{##3}\VANthebibliography}

\usepackage{graphicx}	
\usepackage{amsmath}	
\usepackage{amssymb}	

\usepackage{natbib}
\usepackage{lscape}
\usepackage{rotating}
\usepackage{subfig}
\usepackage{placeins}
\usepackage{hyperref}
\usepackage{txfonts}
\usepackage{comment}
\usepackage{booktabs}
\usepackage[normalem]{ulem} 

\usepackage{xcolor}
\usepackage{xspace}
\usepackage{bm}
\usepackage{longtable}

\makeatletter
\newcommand*\mytabsize{%
   \@setfontsize\mytabsize{1}{1}%
}
\makeatother

\newcommand{\degree}{$^{\circ}$\xspace}
\newcommand{\source}{IGR J17591--2342\xspace}

\newcommand{\msun}{M$_{\odot}$\xspace}

\newcommand{\dl}{\texttt{discline}\xspace}
\newcommand{\sd}{\texttt{shaddisc}\xspace}
\newcommand{\xspec}{\textsc{Xspec}\xspace}

\newcommand{\nicer}{\textit{NICER}\xspace}
\newcommand{\chan}{\textit{Chandra}\xspace}
\newcommand{\xmm}{\textit{XMM-Newton}\xspace}
\newcommand{\nustar}{\textit{NuSTAR}\xspace}
\newcommand{\integral}{\textit{INTEGRAL}\xspace}
\newcommand{\astro}{\textit{ASTROSAT}\xspace}
\newcommand{\accr}{AMXPs\xspace}

\title[Spectral analysis of \source]{Spectral analysis of the AMXP \source during its 2018 outburst}

\author[A. Manca et al.]{
A. Manca,$^{1}$\thanks{E-mail: arianna.manca@dsf.unica.it}
A. F. Gambino,$^{2}$
A. Sanna,$^{1,3}$
G. K. Jaisawal,$^{5}$
T. Di Salvo,$^{2,3,4}$
R. Iaria,$^{2}$
S. M. Mazzola,$^{1}$
\newauthor
A. Marino,$^{8,9}$
A. Anitra,$^{2}$
E. Bozzo,$^{7}$
A. Riggio,$^{1,4}$
and L. Burderi,$^{1,3,4}$ \\
$^{1}$Dipartimento di Fisica, Universit\`a degli Studi di Cagliari, SP Monserrato-Sestu, KM 0.7, Monserrato, 09042 Italy \\
$^{2}$Universit\`a degli Studi di Palermo, Dipartimento di Fisica e Chimica - Emilio Segr\`e, via Archirafi 36 - 90123 Palermo, Italy\\
$^{3}$INFN, Sezione di Cagliari, Cittadella Universitaria, 09042 Monserrato, CA, Italy \\
$^{4}$INAF - Osservatorio Astronomico di Cagliari, via della Scienza 5, 09047 Selargius (CA), Italy \\
$^{5}$National Space Institute, Technical University of Denmark, Elektrovej 327-328, 2800 Lyngby, Denmark \\
$^{6}$Istituto Nazionale di Astrofisica, IASF Palermo, Via U. La Malfa 153, I-90146 Palermo, Italy \\
$^{7}$ISDC, Department of Astronomy, University of Geneva, Chemin d'\'Ecogia 16, 1290 Versoix, Switzerland \\
$^{8}$Institute of Space Sciences (ICE, CSIC), Campus UAB, Carrer de Can Magrans s/n, E-08193 Barcelona, Spain \\
$^{9}$Institut d'Estudis Espacials de Catalunya (IEEC), E-08034 Barcelona, Spain  \\
}

\date{Accepted XXX. Received YYY; in original form ZZZ}

\pubyear{2015}

\begin{document}
\label{firstpage}
\pagerange{\pageref{firstpage}--\pageref{lastpage}}
\maketitle

\begin{abstract}

The Accreting Millisecond X-ray Pulsar \source is a LMXB system that went in outburst on August 2018 and it was monitored by the \nicer observatory and partially by other facilities. We aim to study how the spectral emission of this source evolved during the outburst, by exploiting the whole X-ray data repository of simultaneous observations. The continuum emission of the combined broad-band spectra is on average well described by an absorbed Comptonisation component scattering black-body-distributed photons peaking at (0.8$\pm$0.5) keV, by a moderately optically thick corona ($\tau$=2.3$\pm$0.5) with temperature of (34$\pm$9) keV. 
A black-body component with temperature and radial size of (0.8$\pm$0.2) keV and (3.3$\pm$1.5) km respectively is required by some of the spectra and suggests that part of the central emission, possibly a fraction of the neutron star surface, is not efficiently scattered by the corona. The continuum at low energies is characterised by significant residuals suggesting the presence of an absorption edge of \ion{O}{viii} and of emission lines of \ion{Ne}{ix} ions. Moreover, broad \ion{Fe}{i} and \ion{Fe}{xxv} K$\alpha$ emission lines are detected at different times of the outburst, suggesting the presence of reflection in the system.
\end{abstract}




\section{Introduction}

Accreting Millisecond X-ray Pulsars (\accr) are Low Mass X-ray binary systems (LMXBs) hosting neutron stars (NS) that show coherent pulsations with periods lower than 10 ms. Neutron Stars in these systems are characterised by low magnetic fields, generally between $10^{8}$--$10^{9}$ G \citep[see, e.g.,][]{DiSalvo_AMXP}. Their relatively small spin periods are  now established to be a direct consequence of the mass transfer occurring via Roche lobe overflow from a low mass (<~1~\msun) companion star onto a slow rotating NS. This property makes \accr the progenitors of the rotation-powered millisecond pulsars emitting on a large fraction of the electomagnetic spectrum, i.e. from the radio to the gamma-ray band \citep[][]{Alpar_82}.  
AMXPs usually show sporadic outbursts during which the X-ray luminosity can attain values between $10^{36}$ and $10^{37}$ erg/s.  
So far, 24 sources are included in this subclass of objects \cite[see][]{Bult2022, Ng_21}. Even though the spectral evolution of these sources was rarely monitored in detail during the whole outburst, \accr are generally observed in hard spectral states with no hard to soft transitions. For this reason \accr are usually referred as hard X-ray transients \citep[][]{DiSalvo_AMXP}. 
 
The AMXP system \source was discovered in outburst by the INTernational Gamma-Ray Astrophysics Laboratory (\integral) on August 2018 \citep[][]{Ducci_18}. \cite{Nowak_19} inferred the celestial coordinates of the source from a pointed \chan/HETG observation, i.e. 17$^{h}$ 59$^{m}$ 02.83$^{s}$, -23\degree43' 10.2" (J2000) with an error box of 0.6" representing the 90\% confidence level of \chan positional accuracy. Observations performed by the Australia Telescope Compact Array (\textit{ATCA}) detected the radio counterpart of \source providing an improvement on the previous coordinates that resulted to be 17$^{h}$ 59$^{m}$ 02.86$^{s}$ $\pm$ 0.04, -23\degree43' 08.3"~$\pm$ 0.1" (J2000) \citep[][]{Russell_18}.

Using \nicer and \nustar observations, the timing analysis performed by \cite{Sanna_18} allowed to constrain coherent pulsations at a period of 1.9 ms and an orbital period of 8.8 hrs from the Doppler modulation. Moreover, the X-ray data show a peculiar spin-down behaviour that is compatible with a magnetically threaded disc \citep[][]{Sanna_20}. The results reported by \cite{Sanna_18} led to discuss a scenario in which the companion star is a late spectral type star having an age between 8--12 Gyrs and with a mass between 0.85--0.92 \msun, assuming a mass of 1.4 \msun for the NS. This scenario is consistent with a value of the inclination angle between 28\degree--30\degree.

Broad-band spectral analysis was performed on this source, but considering an average spectrum obtained by pointed observations at different times of the outburst.
\cite{Sanna_18} analysed a combined averaged spectrum consisting in Swift, \integral/ISGRI and \nustar FPMA and FPMB data. They modelled the spectrum using an absorbed soft black-body plus a Comptonisation component. Their work reports a value of the hydrogen column density $N_H$ of (3.6$\pm$1.1)$\times 10^{22}$ cm$^{-2}$, and a Comptonisation component that is characterised by a photon index $\Gamma$ of about 1.8, a temperature $kT_e$ of the corona of 22$^{+4}_{-3}$ keV, and a temperature of the seed photons (assuming a black-body spectrum) of $kT_{seed} = 0.79 \pm 0.09$ keV. The soft black-body direct emission is characterised by a temperature that is compatible with that of the seed photons and in line with the emission from a region of a few kilometers. These authors also detected a weak Gaussian line centred at an energy compatible with the iron K$\alpha$ complex, and interpreted as a possible signature of disc reflection.

\cite{Nowak_19} modelled a 1--9 keV \chan-\nicer spectrum of the source, taken at about 58353 MJD, using a model consisting in an absorbed black-body and Comptonisation component. They inferred a value of $N_H$ of (2.9$\pm$0.5)$\times 10^{22}$ cm$^{-2}$ and a black-body temperature of 0.06 keV, by fixing the $kT{_e}$ value to that observed by \cite{Sanna_18}. In addition, based on a \ion{Si}{xiii} absorption line detected in the \chan spectrum, they propose the presence of an outflow with a velocity of about 2800 km s$^{-1}$ in the system. They also found evidences of possible Ca lines in the HETGS spectra and hypothesized that the NS could be formed via accretion induced collapse of a white dwarf in a rare, calcium-rich Type Ib supernova explosion.

\cite{Kuiper_20} studied the broad-band 0.3--150 keV averaged spectrum of the source using data from \nustar FPMA and FPMB modules, from \xmm RGS, Epic-pn and Epic-MOS2 instruments and from \integral/ISGRI. They modelled the spectrum with an absorbed Comptonisation component finding a value of $N_H$ of (2.09$\pm$0.05)$\times 10^{22}$ cm$^{-2}$ and a Comptonisation component in which the temperature of the seed photons is about 0.64 keV, the coronal temperature $kT_e$ is (38.8$\pm$1.2) keV and in which the optical depth of the corona is $\tau= 1.59\pm0.04$. They excluded the presence of a local absorber into the  system, due to the fact that their estimation of $N_H$ results in line with the one expected according to the total Galactic absorption (i.e. 2.2$\times 10^{22}$ cm$^{-2}$) resulting from the optical reddening maps \citep{Russell_18}. They also found evidences of an emission line in the Iron K-$\alpha$ region. However, they considered the detection of this line spurious and related to blending of lines from different Fe ionisation stages, or alternatively to uncertainties in the XMM Epic-pn response \citep[][]{Kuiper_20}. In the same work, these authors estimated an upper limit on the distance to the source of $d=(7.6\pm0.7)$ kpc according to the analysis of a burst showing clues of Photospheric Radius Expansion (PRE) in one \integral/JEM-X observation.

In this work we perform a spectral study of \source based on the entire \nicer data set and including a large sample of the available observations in the X-ray archive, with the aim of studying its spectral evolution in detail during the whole outburst. The paper is structured as follows: in  \autoref{sec:reduction} we describe the data selection and reduction, in \autoref{sec:analysis} we report the data analysis, in \autoref{sec:discussion} we discuss the obtained results, and in \autoref{sec:conclusion} we summarize these results advancing some conclusions.

\section{Observations and data reduction}
\label{sec:reduction}

\begin{figure*}
\centering
\includegraphics[angle=0, scale=0.8]{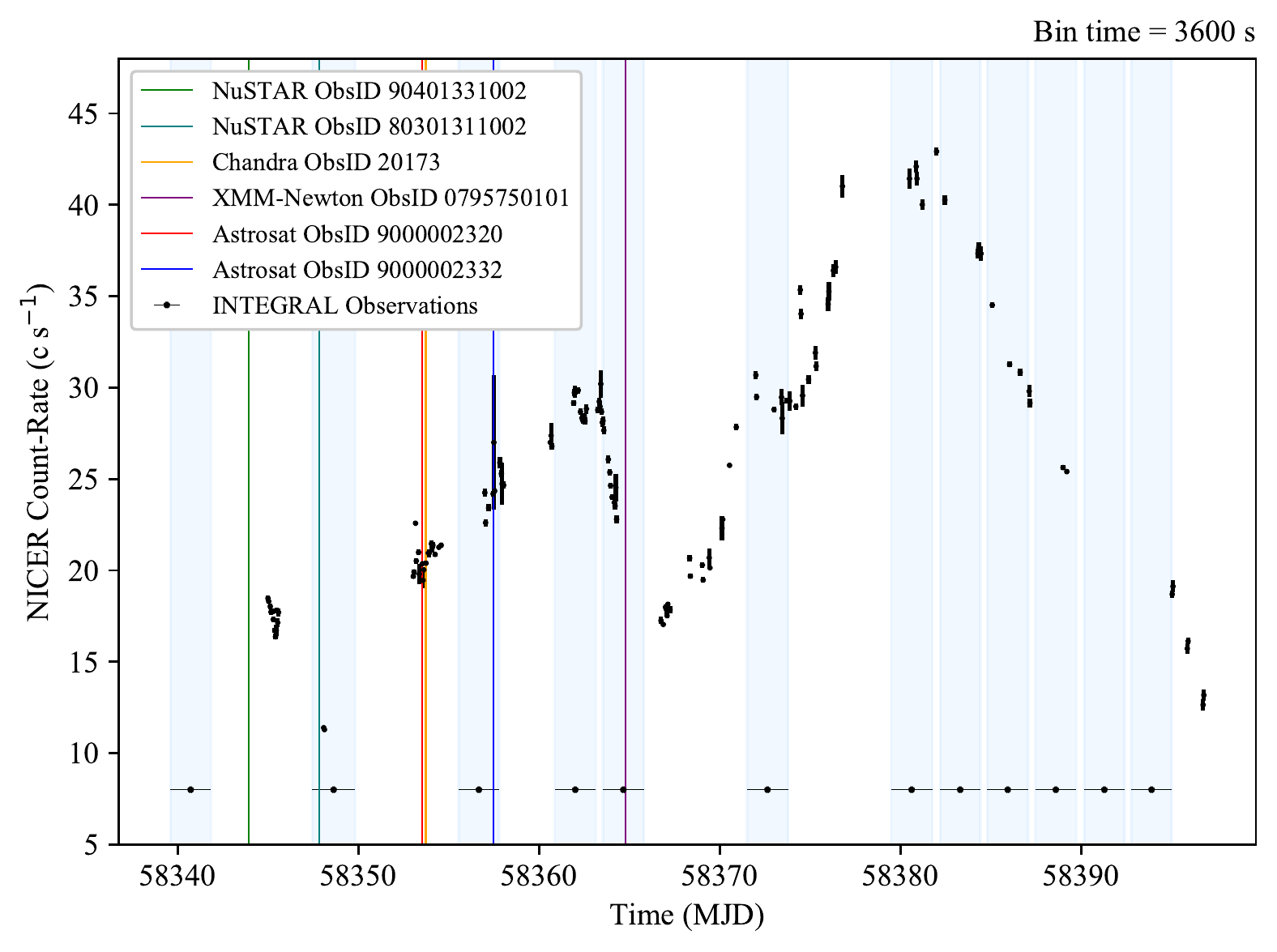}
\caption{NICER light curve of \source during the outburst of 2018. Superimposed we show the mid-observation times at which other observatories pointed the source and the shaded areas representing the time windows of the \integral observations.}
\label{fig:outburst_profile}
 \end{figure*}

 \begin{figure}
\centering
\includegraphics[angle=0, scale=1.0]{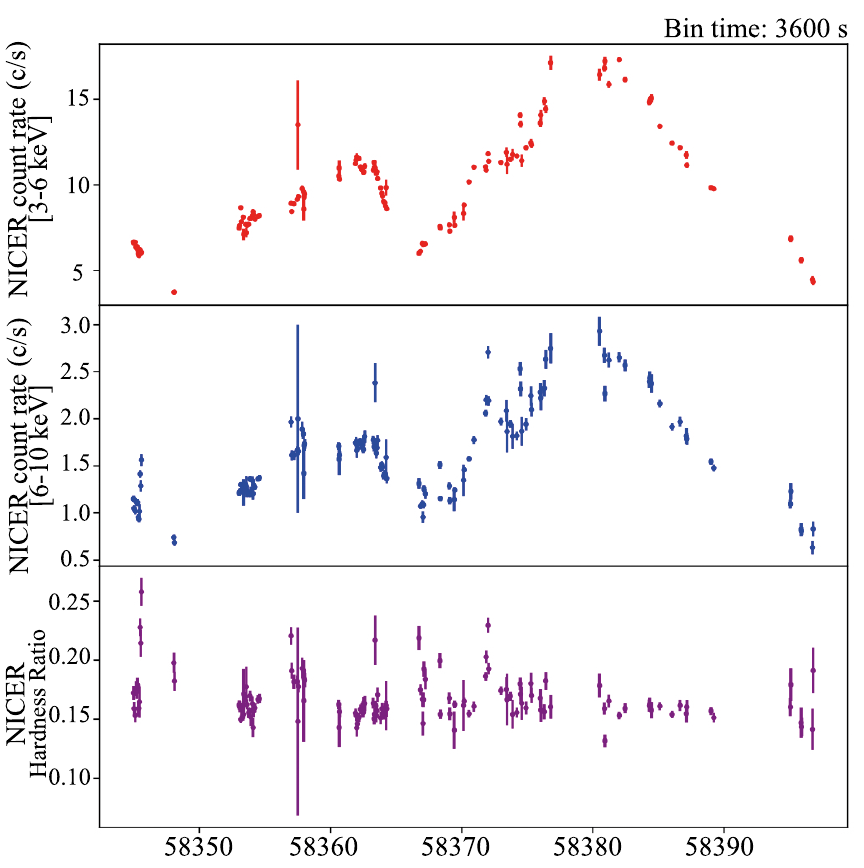}
\caption{The 3--6 keV (upper panel) and 6--10 keV (middle panel) \nicer light curve of \source during the outburst of 2018. In the bottom panel we report the hardness ratio evaluated between these two energy bands. }
\label{fig:HR}
 \end{figure}
A complete follow up of the outburst of \source was performed by the X-ray timing instrument (XTI, \citealt{Gendreau2012,  Gendreau2016}) on board the Neutron star Interior Composition Explorer (\nicer) mission, hosted onto the International Space Station (ISS). This instrument collected data of the source between 2018-08-14T23:59:42 (ObsID 1200310101) and 2018-10-17T05:32:22 (ObsID 1200310139). 
We reduced the \nicer data using the {\tt nicerl2}\footnote{\url{https://heasarc.gsfc.nasa.gov/docs/nicer/analysis_threads/nicerl2/}} script available under \textsc{HEAsoft} version 6.28. We processed the data using the latest gain and calibration database files of version 20200722. We applied standard filtering criteria based on elevation angle from the Earth limb, pointing offset, offset from the bright Earth, and South Atlantic Anomaly region in our study. Following the above criteria, we selected the Good Time Intervals using the {\tt nimaketime} task. We extracted the source spectrum using the filtered clean event file in the \textsc{xselect} environment, while we obtained the background spectra corresponding to each observation using the {\tt nibackgen3C50}\footnote{\url{https://heasarc.gsfc.nasa.gov/docs/nicer/toolsnicer_bkg_est_tools.html}} tool \citep[][]{Remillard2021}. For the spectral analysis, we considered the response matrix and ancillary response file of version 20200722. We added a systematic error of 1\% to the obtained source spectra, as suggested by the \nicer team \citep[see, e.g.,][]{Jaisawal_19}.

From \autoref{fig:outburst_profile}, it is possible to follow the complete light curve of the source during the outburst as observed by \nicer in the band 0.2--10 keV. The first observation of the source shows an average count-rate of about 17 c/s that increased up to a first peak of 29 c/s at 58362 MJD. After this first maximum, the count-rate decreased until it reached a value of about 17 c/s at 58367 MJD, when it started to increase again up to the absolute maximum of about 42 c/s reached at 58380 MJD. Later on, the source started to decrease its luminosity down to a count-rate of 14 c/s, reached on 58397 MJD. In addition, as highlighted by the hardness ratio reported in \autoref{fig:HR} and evaluated between the energy band 3--6 keV and 6--10 keV, the spectral state of \source maintains quite constant, suggesting that no significant spectral changes occurred during the outburst, in agreement with what usually observed for \accr \citep[][]{DiSalvo_AMXP}.

The outburst of \source was also observed by \integral between revolutions 1986 and 2009. In addition, further single pointed observation of the Nuclear Spectroscopic Telescope Array (\nustar), \chan, \xmm and \astro were performed during the first part of the outburst.

\nustar \citep[][]{Harrison_13} observed \source on 2018-08-13T22:36:09 (ObsID 90401331002) and on 2018-08-17T20:01:09 (ObsID 80301311002). The data have been reprocessed with the {\sc nupipeline} routine while the source and background spectra were extrapolated with the {\sc nuproducts} pipeline, using circular extraction regions of 60" of radius centred at the coordinates of the source and far away from the source, respectively. The spectra were modelled in the 2--80~keV energy band.

\chan observed \source on 2018-08-23T17:40:05 for 20 ks using the High Energy Transmission Grating spectrometer \citep[HETG,][]{Canizares_05}.
The events were reprocessed using the official Chandra analysis environment {\sc CIAO} v. 4.13 with calibration files updated to the version 4.9.4. The data were reprocessed with the {\sc chandra\_repro} pipeline and the first order High-Energy Grating (HEG) and Medium-Energy Grating (MEG) spectra were combined using the {\sc combine\_spectra} tool. The obtained spectra were grouped to have at least 25 counts per energy bin and analysed in the 1.2--7.8~keV energy band.

The \xmm observatory observed the source with both the Reflection Grating Spectrometer \citep[RGS,][]{Herder_01} and the European Imaging cameras (EPIC) on 2018-09-03T18:44:55 during a time window in which the count-rate of the source momentarily decreased, as it can be inferred from \autoref{fig:outburst_profile}. The science data reduction was performed using the \xmm Science Analysis System (SAS) version 19.0.0.
The (imaging) EPIC-pn instrument \citep[][]{Struder_01} operated in Timing Mode, while the MOS-1 and MOS-2 cameras \citep[][]{Turner_01}, acquired data in Small window and Timing uncompressed modes, respectively.
The Epic-pn spectra were extracted from the RAWX coordinates between 31--44, while the background spectrum was obtained from RAWX coordinates centred between 5 and 15. The MOS 1 data were collected in Small window mode and, as already noticed by \cite{Kuiper_20}, are extremely corrupted by pile-up effects and for this reason they are not considered in our work. On the contrary, MOS 2 data are not affected by pile-up, since the instrument was operating in Timing uncompressed mode, which allows to collect data up to 35 mCrab without severe pile-up distortions. The source and background spectra for the MOS2 were extrapolated in the RAWX range 290--320, while the background spectrum was extracted by a box having a width and a height of 8432.64 and 4285.44 in physical coordinates, respectively. The analysed energy range is 2.2--10~keV.
The RGS operated in the standard Spectroscopy HighEventRate mode with Single Event Selection. No soft proton background flares were detected during the observation. Then, we combined the first order data of RGS1 and RGS2 using the {\sc rgscombine} tool included in the SAS package, and grouping the obtained spectrum to have at least 25 counts per energy bin. However, the data counts are considerably low for these data for energies below 1.2 keV, and for this reason we preferred not to consider these data for our analysis, since the energy range above 1 keV is well covered by the rest of the considered space missions.

The \astro mission observed the source with the Large Area X-ray Proportional Counter (LAXPC) instrument \citep[][]{Yadav_16,Antia_17,Agrawal_17} on 2018-08-23T01:10:15 (ObsID 9000002320) and on 2018-08-27T00:00:00 (ObsID 9000002332) for a net exposure of 30 ks and 37.7 ks, respectively. A thermonuclear type I X-ray burst occurred during the ObsID 9000002320 and has been removed before extracting the persistent spectrum of the source. For both the observations, the LAXPC 30 was not working. On the other hand, the LAXPC 10 was active on a low voltage gain setting\footnote{\url{http://astrosat-ssc.iucaa.in/}}. Since the source is faint, we extracted spectra from the top layer of each detector to avoid unnecessary background. However, the LAXPC 10 spectra appear to be affected by the background and gain change, and for this reason they were not included in the analysis.
In addition, the LAXPC 20 spectrum extracted for ObsID 9000002320 shows a significant mismatch with respect to the spectra of the closest available observations from other missions, possibly introduced by a bad calibration. For this reason, this observation was also excluded from the analysis.
A systematic error of 2\% has been added to the LAXPC 20 spectrum of ObsID 9000002332 \citep[see, e.g.,][]{Misra2017}, which was also grouped to have at least 25 counts per energy bin. The analysis was conducted in the 4--17~keV energy range.

\integral observed the outburst from the source during the period spanning from 58340 MJD to 58405 MJD, corresponding to satellite revolutions from 1986 to 2016. We considered data from the two JEM-X units \citep[][]{Lund_03} and from IBIS/ISGRI \citep[][]{Umbertini_03,Laurent_03}.
We analysed the ISGRI, JEM-X 1 and JEM-X2 data with the Offline Scientific Analysis (OSA) software version 11.0 distributed by the ISDC \citep[][]{Courvoisier_03}. Different source spectra were initially extracted for the two JEM-X and IBIS/ISGRI from each revolution. A grouping of 16 bins has been used for the JEM-X (IBIS/ISGRI) spectra extraction, following the standard practice for similar sources. We excluded the science window (SCW) 50 in revolution 2001 due to the presence of a type I X-ray burst \citep[see also][]{Kuiper_20}. The JEM-X spectra were analysed in the 3--20~keV energy range, while the IBIS/ISGRI spectra were analysed in the 25--200~keV range.

\begin{figure*}
\centering
\includegraphics[angle=0, scale=0.95]{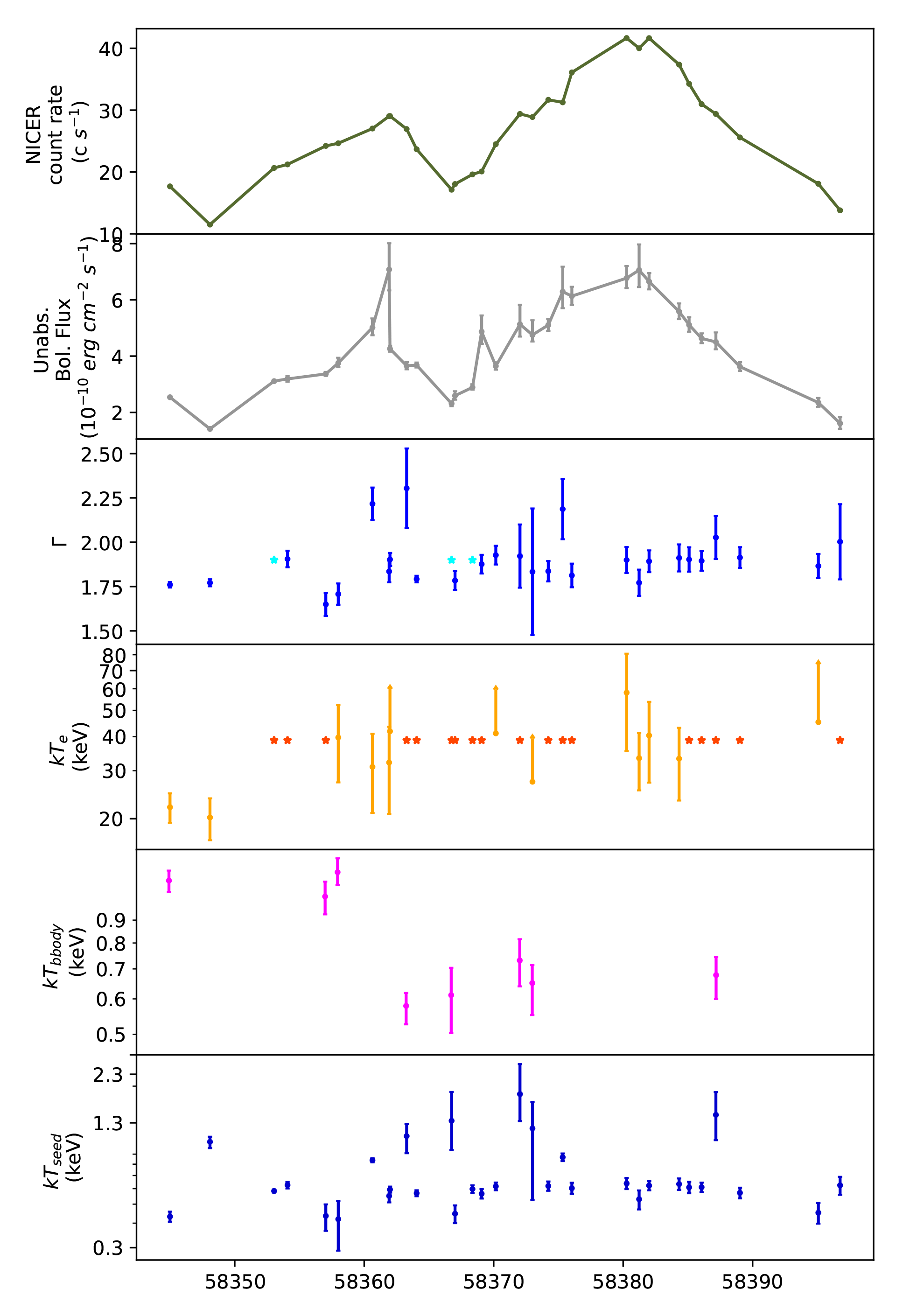}
\caption{Evolution of the main spectral parameters describing the continuum emission of \source as a function of time. The associated errors are reported at a level of statistical confidence of 90\%. In the upper panel the mean count-rate collected during each \nicer observation is reported. In each panel, the star-shaped points represent fixed values for the parameters.}
\label{fig:par_evolution}
 \end{figure*}

\begin{figure}
\centering
\includegraphics[angle=0, scale=0.5]{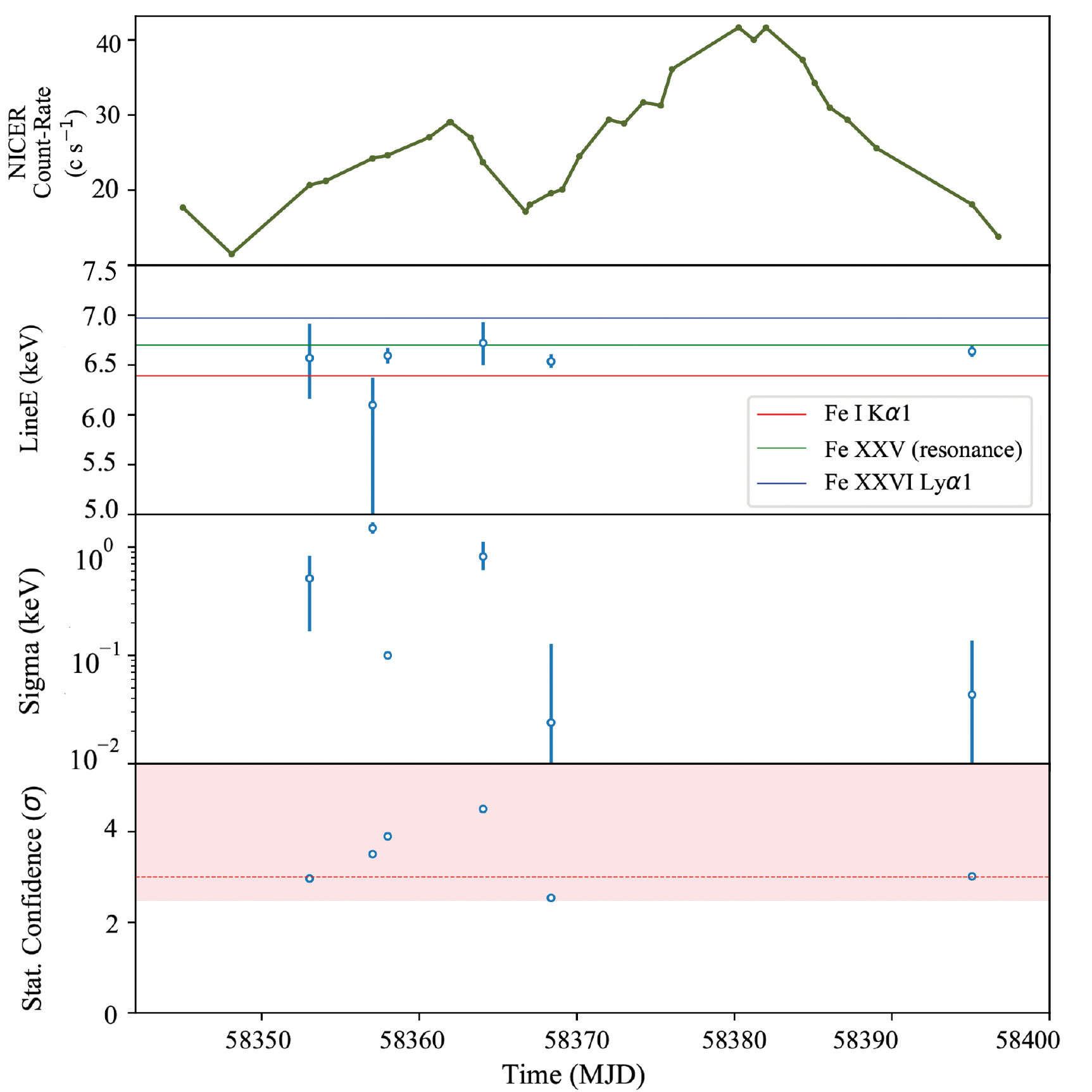}
\caption{Evolution of the iron line spectral parameters and statistical significance of the observed feature. In the plot of the line centroids (second plot from the top) we also reported the rest-frame energy of the \ion{Fe}{i} K$\alpha1$ , \ion{Fe}{xxv} (resonance), and \ion{Fe}{xxvi} Ly$\alpha1$ transitions, as reference. In the bottom panel, we highlighted in pink the detection area considering as lower limit $\sigma$=2.5 for a weak detection. The dotted line represents the 3$\sigma$ detection threshold.}
\label{fig:Fe_evolution}
 \end{figure}

\begin{figure}
\centering
\includegraphics[scale=0.35]{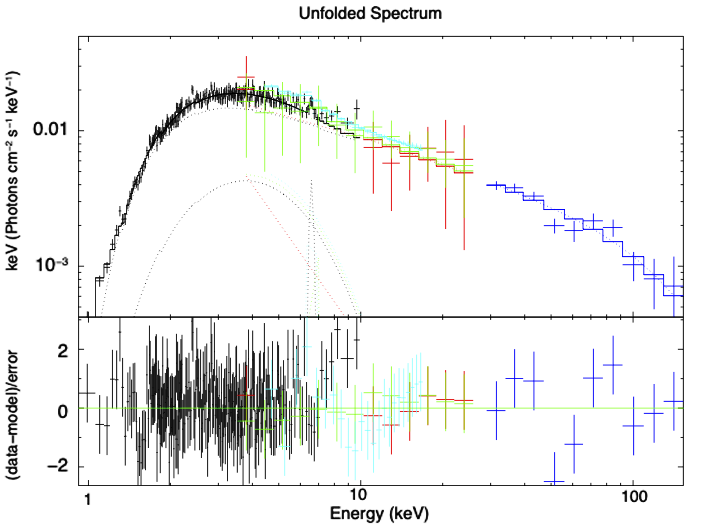}
\caption{Best fit model and associated residuals obtained for the broad-band spectrum associated to the \nicer ObsID 1200310106 (see \autoref{tab:continuum} and \autoref{tab:lines}). In black the \nicer spectrum, in red, green and blue the \integral JEM-X1, JEM-X2 and ISGRI spectra of revolution 1992 respectively, and in cyan the \astro LAXPC20 spectrum of ObsID 9000002332. }
\label{fig:unfolded_spectrum_106}
 \end{figure}

\section{Spectral Analysis}
\label{sec:analysis}

The spectral analysis was entirely performed using the X-ray spectral fitting package \xspec v. 12.11.1c \citep[][]{Arnaud96}.
We chose to fit as many spectra as possible in order to follow the evolution of the spectral parameters during the outburst with the highest possible temporal resolution guaranteed by the available data. For this reason, we paired each \nicer spectrum with any quasi-simultaneaous, i.e. taken within two days, data-set from the other observatories, when possible. 
As widely discussed in \cite{Sanna_18} and \cite{Kuiper_20}, after some preliminary tests on the available data, we noted that the model which best describes the continuum was an absorbed Comptonisation model. For this reason, we adopted this model in the following analysis. Some of the performed fits, however, revealed that in some cases a soft excess remained in the residuals and this could be corrected with the addition of a black-body emission from the source, as already reported in \cite{Sanna_18}. 
We used the chemical abundances of \cite{wilms} and the cross sections reported in \cite{Verner}.
We modelled the continuum spectral emission of \source adopting the Tuebingen-Boulder ISM absorption model {\sc Tbabs}, a black-body component {\sc bbodyrad}, and the thermal Comptonisation component {\sc nthcomp}. 
The {\sc nthcomp} model is described by the asymptotic power-law photon index $\Gamma$, the electron temperature $kT_e$ of the hot electron corona, the seed photon temperature $kT_{seed}$, the $inp\_type$ parameter, which can assume values 0 or 1 for considering black-body or disc-black-body distributions for the seed photons, respectively, and the $redshift$ parameter. We assumed that the seed photons are distributed accordingly to a black-body law by fixing the $inp\_type$ parameter to 0. Moreover, we adopted a value of redshift equal to zero for all the subsequent analysis. The \nicer spectra were fitted in the energy range 0.6--10 keV. On the other hand, the \nustar FPMA and FPMB spectra of ObsID 80301311002 were fitted in the range 2--60 keV, while those of ObsID 90401331002 in the range 2--80 keV. 
The \xmm spectra of ObsID 0795750101 were fitted in the range 3--10 keV and 2.2--10 keV for the EPIC-pn and EPIC-MOS2 spectra, respectively. 
The \chan MEG and HEG spectra of ObsID 20173 were fitted in the range 1.2--6.8 keV and 1.4--7.8 keV, respectively, while the \astro/LAXPC 20 spectrum of ObsID 9000002332 was fitted in the energy range 4--17 keV.

We inspected with \textsc{xspec} all \integral data of the source, but for our broad-band analysis we finally only made use of the spectra extracted during revolutions from 1992 and 2006 because they are more reasonably close in time to the available \nicer data. During revolutions from 1986 to 1989, the source was caught by \integral during the earliest stages of the outburst evolution and the \integral data were characterised by a number of counts too low to perform any meaningful spectral analysis. A similar conclusion applies to the data collected during revolutions 2008--2016, corresponding to the rapid decay of the source flux toward the end of the outburst. Depending on the source flux, we limited our spectral analysis for the JEM-X data roughly in the interval 3--20 keV. During revolutions 1989, 1994, 1995, 2002, 2004 and 2006, the JEM-X data did not have a large number of counts, in order to be used in the spectral analysis and thus we do not mention these data any longer in the following sections. IBIS/ISGRI data were used, following the OSA 11.0 recommendations, from an energy of 25 keV up to roughly 200 keV, above which the source fell below the detection threshold of the instrument. The higher energy bound of the IBIS/ISGRI spectrum varies across the different revolutions considered depending on the flux of the source but remains in all cases well above 100 keV.

The \nicer spectra, in combination with those of other missions, generally ensured a good energy coverage over a large energy range. However, some of the \nicer spectra could not be combined with other spectra, keeping hard to constrain the high-energy cut-off $kT_{e}$ for the Comptonisation component. In all these cases, this parameter was fixed to the value obtained by \cite{Kuiper_20} (i.e. $kT_{e}=38.8 \;keV$). In \autoref{fig:par_evolution} we show the evolution of the spectral parameters obtained from the best fit model of each observation as a function of time. In all the plots, the star-shaped points indicate parameters that were frozen in the fit. This plot shows how the $kT_{e}$ parameter has been constrained for many observations, with an average value obtained during the whole outburst of about 34 keV, which is in line with that inferred by \cite{Kuiper_20}. 
Some of the \nicer spectra showed the presence of residuals in absorption at about 0.87 keV, which are consistent with the presence of an \ion{O}{viii} absorption edge. 
To fit the low energy residuals, as a first step, we tried to replace the {\sc Tbabs} ISM absorption component with the {\sc Tbfeo} and the {\sc Tbvarabs} components that take into account variable abundances for the chemical species in the ISM. However, even if both these models return perfectly compatible values for the $N_H$, they are unable to fit the observed residuals. For this reason we manually described the observed feature by recurring to the {\sc edge} component in \xspec. The energy of the edge was kept frozen for all the fits. 
Other localised residuals characterise some of the \nicer spectra at about 0.922 keV, in line with the presence of \ion{Ne}{ix} ions. This feature was modelled with a {\tt Gaussian} component in which the energy centroid was fixed to the rest-frame energy of the aforementioned ion. Moreover, the spectral resolution of the \nicer/XTI instrument around 1 keV was not sufficient to constrain the width of the modelled line. For this reason, we fixed the parameter $sigma$ to 0.085 keV (i.e. the \nicer/XTI spectral resolution at 1 keV). On the other hand, the normalisation of the Gaussian component was left free to vary.
Firstly, the neutral column density N$_{H}$ in the {\sc Tbabs} component was left free to vary during the whole duration of the outburst, but we noticed a correlation with the kT$_{seed}$ parameter of the Comptonisation component, with a consequent scattering of the N$_H$ component. Therefore, we decided to keep this parameter frozen at the value of (2.09$\pm$0.05)$\times 10^{22}$ cm$^{-2}$, in accordance with \cite{Kuiper_20} for the rest of the analysis.

The Comptonisation component is on average characterised by stable values of $\Gamma$ and $kT_e$, equal to 1.9, with a standard deviation of about 0.2, and 34 keV with a standard deviation of about 9 keV, respectively. On the other hand, during the outburst the temperature of the seed photons, assumed to be injected with a black-body-like spectrum, shows a variation in accordance with the advance of the outburst. The mean value of this parameter during the outburst was of 0.8 keV with an associated standard deviation of 0.5 keV.

The {\tt bbodyrad} component appears to be required by the fit only in some of the observations. We plotted the resulting black-body temperature $kT$ in the fifth panel from the top of \autoref{fig:par_evolution}. This parameter shows an almost constant trend during the outburst with a mean value of 0.8 keV and a standard deviation of about 0.2 keV. The associated errors at 90\% of confidence level strongly suffer from the statistics of the fitted spectra. However, for all the reported black-body components the significance of its detection results to be higher than 3$\sigma$ (the magenta points have a significance of more than 5$\sigma$, whilst the dark violet ones have a significance between 3$\sigma$ and 4$\sigma$).

Some of the observed combined spectra also showed significant residuals in the energy range at which the \ion{Fe}{xxv} line is expected, i.e at about 6.5 keV. Once the residuals have been detected, we tried to model them with a {\tt Gaussian} component. For each of these observations we evaluated the significance of this local feature, by evaluating how much the normalization of this component deviates from the continuum in units of $\sigma$. In the lower panel of \autoref{fig:Fe_evolution}, we report the significance of the observed iron line. In the upper and medium panels of the same figure, we show how the parameters of this feature evolve in time during the outburst. In particular, we can observe that, with the exception of the second point for which we have an upper limit on the energy centroid, the rest of the lines have been well constrained, even though it was not possible to obtain lower limits for the associated width. However, these observations suggest that the line is marginally detected and that results to be in agreement with a \ion{Fe}{i} K$\alpha1$ line the first phase of the burst. After this phase, the line starts to be consistent with the presence of more ionised species as \ion{Fe}{xxv}.
Since this line tends to be generally broad, when its width could be constrained, we tried to fit this line also by using the Relativistic model \dl \citep[][]{Fabian89} or the most recent \sd \citep[][]{LaPlaca20}. These models, however, with the exception of the centroid energy, returned totally unconstrained spectral parameters and generally not physically reasonable due to the low number of counts on the line profile.

As an example, we show the best fit model obtained for the broad-band spectrum associated to the \nicer ObsID 1200310106 in \autoref{fig:unfolded_spectrum_106}.

\section{Discussion}
\label{sec:discussion}

\subsection{Spectral evolution}
We report hereon the spectral analysis of the whole set of \nicer observations of the outburst of \source occurred in 2018, integrating these observations with all the available quasi-simultaneous observations stored in the X-ray data archive. The \nicer light curve during the outburst is characterised by a double peak. The spectral model adopted to fit the spectra of the source suggests that during the first peak, occurring at 58362 MJD, the source showed an unabsorbed bolometric flux of (4.6$\pm$0.2)$\times10^{-10}$~erg~cm$^{-2}$~s$^{-1}$ in the range 0.1--100 keV, which increases up to 7.2$^{+0.7}_{-0.5}\times10^{-10}$~erg~cm$^{-2}$~s$^{-1}$ in the same energy band at 58380 MJD, when the second peak of the outburst occurred. The flux errors are reported at 3$\sigma$ c.l., as derived from the \texttt{cflux} convolution model.

Comptonisation is the physical process for which we observe the major contribution in terms of flux. Actually, this component, on average, contributes to the total unabsorbed flux for 95\% in the case of \source. According to the obtained results, the Comptonisation component appears to be characterised by an electron corona with a temperature $kT_e$ of about 34~keV, which is in line with the value of 38.8~keV reported by \cite{Kuiper_20}. Unfortunately, due to the lack of coverage at the higher energies in some points of the outburst, we were unable to constrain this parameter for all the observations performed by \nicer. However, we constrained this parameter in 29\% of the \nicer observations, inferring that at the beginning of the outburst the cloud was characterised by a slightly lower temperature of about 24~keV.

The photons that are Comptonised in this corona appear to be mainly with a temperature of about 0.8~keV, with the tendency to slightly increase during the peaks of the outburst, always, however, at the limit of statistical compatibility at the 90\% confidence level. This could be due to episodes of increased heating of the neutron star surface/boundary layer, plausibly caused by occasional rises in the mass-accretion rate. The asymptotic power law index $\Gamma$, on the contrary, remains almost stable between 1.7 and 2.1.

\begin{figure}
\centering
\includegraphics[angle=0, scale=0.4]{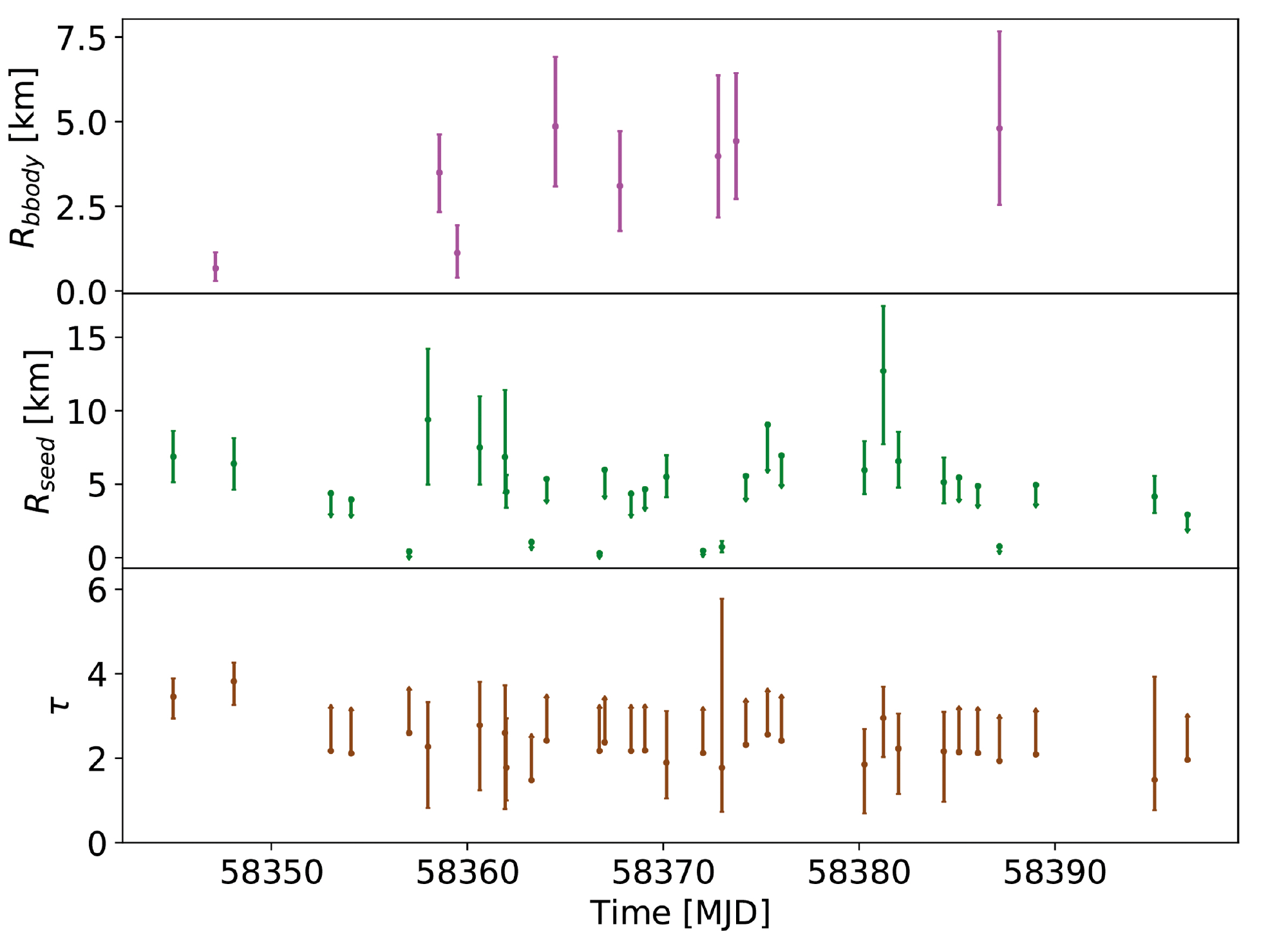}
\caption{Trend of the equivalent radius of emission of the black-body component (upper panel) and of the Comptonisation seed photons (middle panel) as a function of time. In the lower panel, the trend followed by the optical depth of the Comptonising electron cloud is also reported. The arrows indicate upper limits on the values of the parameters.}
\label{fig:radii}
\end{figure}

The black-body component has been tested for each spectrum and only 26\% of the best fit models require this component at a level of confidence that is greater than 3$\sigma$. The equivalent radius of emission of the black-body component has been obtained from the normalization as R$_{bb}$=$N_{bb}d^{2}_{10}$, where N$_{bb}$ is the normalization of the \texttt{bbodyrad} component and $d^{2}_{10}$ is the distance to the source in units of 10 kpc. Assuming the distance to the source obtained in this work (i.e., (7.2$\pm$0.8)~kpc), we obtained the radii reported in \autoref{fig:radii}.

From the observed trend, it is possible to notice how the radius of the emitting black-body is mainly at a value of about $\overline{R_{bb}}$~=~(3.3$\pm$1.5)~km, where the associated error in this case coincides with the standard deviation of the distribution of measurements. This component, peaking mainly at $\overline{kT}$~=~(0.8$\pm$0.2)~keV, could arise from a direct emission by a fraction of the NS surface. It is interesting to notice that the size of this region varies in accordance with the variation of the flux extrapolated in the band 0.1--100 keV during the outburst. In particular, for lower values of the flux, it appears smaller with respect to the case in which the flux is higher. One possible explanation could be represented by the fact that when the mass accretion rate increases, the region responsible for the emission becomes larger, reaching a maximum in proximity of the two peaks of the outburst, where the values are consistent with a radius of about 5~km, i.e half of the NS size, assuming a NS of 10 km of radius. Then, we tried to test if the direct observation of the black-body emission from the NS might be related to a low optical depth of the hot corona, using the relation of \cite{Zdziarski_1996}:

\begin{equation}
\label{eq:Gamma_tau}
\Gamma=\left[ \dfrac{9}{4}+\dfrac{1}{\tau \left(1+\dfrac{\tau}{3}\right) \left(\dfrac{kT_e}{m_e c^2}\right)}\right]^{1/2} - \dfrac{1}{2},
\end{equation}
with the spectral parameters obtained in the best fit model for each observation. The evolution of this parameter as a function of the observing time is reported in \autoref{fig:radii} in the lower panel. As shown by this trend, the value of the optical depth remains almost stable during the outburst at a value of about $\tau\sim$2.3 and a standard deviation of 0.5, at least for the parameters that were constrained. 
For this reason, it is possible that the corona Comptonises the majority of the photons emitted by the NS surface leaving, however, only a small fraction ($\sim$ 10\%) of them not significantly Comptonised. The latter could contribute to the observed direct black-body component, peaking approximately at the same temperature of the seed photons, as evidenced in the obtained results.

A direct proof of this scenario could be provided by estimating the equivalent radius of the region emitting the seed photons for the Comptonisation, which we expect more or less similar to the radius obtained for the direct black-body emission region. An estimation of this physical parameter can be obtained by using the relation of \cite{Zand_99}, assuming a spherical geometry of the corona,

\begin{equation}
\label{eq:R_seed_sph}
R_0=3\times 10^4 \; d \left( \dfrac{f_{bol}}{1+y} \right)^{1/2} \left(kT_{seed}\right)^{-2},
\end{equation}
where $d$ is the distance to the source in kpc inferred in this work, $f_{bol}$ is the unabsorbed bolometric flux extrapolated from the Comptonisation component in erg~cm$^{-2}$~s$^{-1}$, $kT_{seed}$ is the temperature of the seed photons in keV, and $y=4kT_e max[\tau , \tau ^2]/(m_e c^2)$ is the Compton parameter, in which $kT_e$ is the electron temperature in keV. The radius $R_0$ obtained for each observation is plotted as a function of the observing time in the middle panel of \autoref{fig:radii}. These parameters, at least in the cases in which the statistics of the data allowed to constrain them, are scattered around a mean value of $\overline{R_0}$~=~(5.0$\pm$2.8)~km, which is in agreement with the obtained value for $\overline{R_{bb}}$.

\subsection{Spectral lines}
The spectral continuum of the source is characterised by the presence of several local features. On the one hand, almost the totality of the analysed spectra present an absorption edge at 0.871 keV, which suggests the presence of \ion{O}{viii} ions. Despite the fact that the energy of this feature needed to be fixed due to the complexity of the residuals at the lowest energies in the \nicer spectra, the significance of this feature is always higher than 3$\sigma$ for all the results reported in \autoref{tab:continuum}.
The low energy residuals of some spectra are well fitted by taking into account a \ion{Ne}{ix} emission line for all the observations (see \autoref{tab:lines}). 
In accordance with the statistics of the available spectra and with the spectral resolution of \nicer/XTI, which is not sufficient to constrain the line width, we fixed the parameter $sigma$ of the Gaussian component to a value of 85 eV, which is the spectral resolution of the instrument at 1 keV. Trying to leave this parameter free to vary during the fit, indeed results in a value of width that is of the order of the spectral resolution of \nicer at 1 keV, or slightly higher, even though not constrained. 
This could probably suggest two possible scenarios: on one hand, the lines could be produced approximately in the same region of an accretion disc where they are relativistically broadened due to the proximity to the NS. This scenario, however, is not supported by evidences of any further black-body or multicolour-disc black-body component needed to fit the spectrum and attributable to emission from the disc. On the other hand, this line could be smeared only as a consequence of Compton broadening, or as an effect of the not sufficient spectral resolution of the XTI instrument to resolve a complex of lines at those energies. The test of this latter scenario deserves observations with high spectral resolution instruments. Indeed, the low energy spectrum of the source is completely dominated by the background flux below 1 keV in the case of the pointed observations performed by the \chan/HETG and the two MOS instruments on board the \xmm mission.

About 19\% of the observations show a significant evidence ($>3\sigma$) of Fe emission lines. In \autoref{fig:Fe_evolution}, we show the evolution followed by the parameters describing the line profile, assumed to be Gaussian.
The detection of this line is significant only for the time window including the range spanned by the first peak of the outburst and by part of the rise of flux towards the main peak of the burst, i.e during the periods of higher statistics of the \nicer spectra. Our analysis shows how, in the reported cases, it is possible to constrain the centroid energy of the lines, with the exception of the observations occurred on 58357 MJD and on 58395 MJD, for which we find only an upper limit of 6.3 and 6.7 keV, respectively. The remaining sample of line profiles result to be in line with the rest-frame energy corresponding to the \ion{Fe}{xxv} K$\alpha$ line, usually observed in bright LMXB systems showing evidences of reflection.
The observed sigma for these lines results to be on average of about 0.5 keV and its distribution shows a standard deviation of about 0.5 keV. On the basis of this result, it seems that the line profile is tendentially broad. \cite{Kuiper_20} detected a similar line profile in the average spectrum of the source. The feature they observed is centred at about 6.7 keV and has a width of 0.69 keV, in line with our results. However, they consider this feature to be characterised by values of the centroid energy and sigma that are too far off and too broad. They propose that this feature could be produced by a blend of Fe lines deriving from different ionisation stages, which can not be distinguished individually by the MOS due to its spectral resolution. 
Moreover, they propose that this detection could reflect uncertainties in the \xmm EPIC-pn response for observations taken in timing mode, also because there is no detection for \chan/HETG data, having a higher spectral resolution. For this reason they do not consider the detection as real.
On the contrary, we detected this line at different times during the outburst, including the case occurring on 58353 MJD for which we observe a detection at about 3$\sigma$, and that was obtained by considering a combined spectrum that also includes a \chan/HETG spectrum. In accordance with this evidence, and to the fact that also other spectra during the outburst show the same feature, we conclude that this feature is real.

The poor statistics offered by the available data on the iron line profile, and the lack of detected lines for close observations, does not allow a deeper investigation on the nature of this feature that could be produced by reflection. A simple modelling with a Relativistic component as \dl or \sd returns unconstrained values for the physical parameters describing the line profile. Further observations of the source in outburst, performed with future space missions that will be provided with higher effective area and spectral resolution over a wider energy range, will be fundamental to investigate the reflection component in \source, for a deeper comprehension of the ionisation state of the matter and of the geometry of the system.

\subsection{A new measurement of the distance}
The analysis of PRE type I X-ray bursts is largely used in literature for inferring the distance to the sources, the method being often the only possible way to obtain such an estimate. However, this method suffers from systematic uncertainties \citep[see, e.g.,][and references therein]{Marino_19}. For instance, the flux reached by PRE bursts for the same source generally scatters around a mean value with variations of about 15\% \citep[see, e.g.,][]{Kul_03,Galloway_03,Galloway_08}. Moreover, the value used for the Eddington luminosity of the source may not be accurate without details on, e.g., the composition of the NS atmosphere and and therefore of its opacity. The latest value for the distance of the source is $d=(7.6\pm0.7)$ kpc, obtained by \cite{Kuiper_20} from the analysis of a type-I X-ray burst which was suggested to reach the Eddington limit. However, without strong evidences of PRE, the obtained distance could be an upper limit rather than an actual estimate.  Furthermore, the aforementioned systematic uncertainties that affect the method, demand to try an alternative way to find the distance and eventually confirm the measurement by \cite{Kuiper_20}.

\begin{figure}
\centering
\includegraphics[scale=0.455]{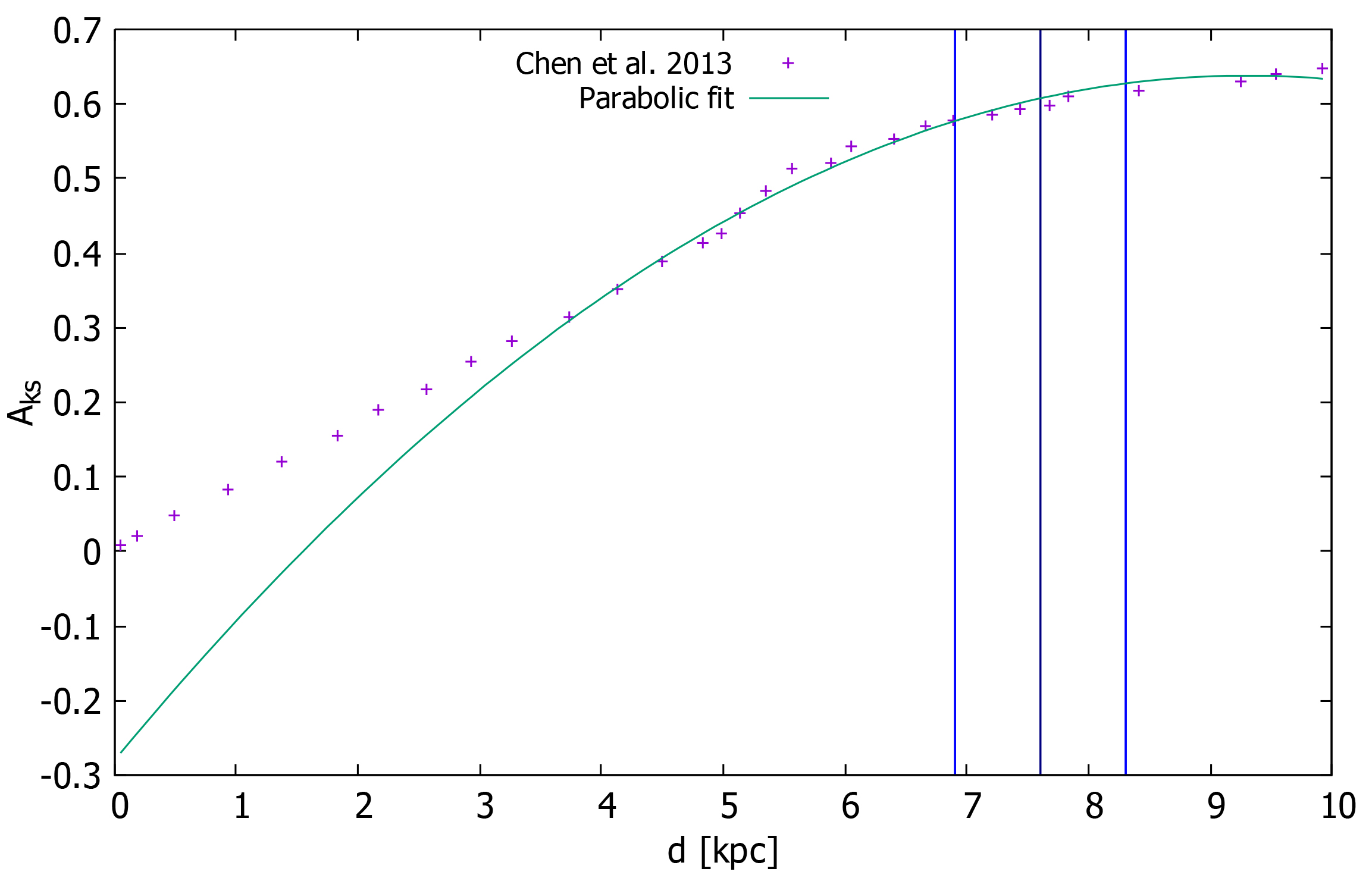}
\caption{Expected profile of the extinction in the K$_{s}$ band, as a function of the distance to the source in the direction of \source. The blue vertical lines indicates the best fit value (navy blue line) of the distance to the source inferred by \citet{Kuiper_20} and its relative error (lighter blue lines). The green line represents the best fit parabolic function that fits the profile in the range 5--10 kpc.}
\label{fig:Ext_profile}
\end{figure}

From the value of the hydrogen column density, we can derive an estimate of the distance of the system to compare with the existing one by \cite{Kuiper_20}, by invoking the 3D extinction map of the radiation in the  K$_{s}$ band for our Galaxy of \citet{Chen2013}. The map provides a profile of the radiation extinction in the direction of the Galactic bulge, as a function of the distance to the source. We used the profile at galactic coordinates $l=0.00$, $b=1.00$, reported in \autoref{fig:Ext_profile}.

The value of N$_H$ is related to the visual extinction of the source radiation A$_{V}$ through the relation of \cite{Guver_09}:

\begin{equation}\label{eq:A_v}
N_{H}=(2.21 \pm 0.09)\times10^{21}A_{V}.
\end{equation}
 
The visual extinction is then related to the extinction of the radiation in the K$_{s}$ band (A$_{K_{S}}$) through the relation of \cite{Nishiyama_08}:
\begin{equation}\label{eq:A_ks}
A_{K_{S}}=(0.062\pm0.005)\;A_{V}\;{\rm mag}.
\end{equation}

We fitted the profile of \autoref{fig:Ext_profile} with a parabolic function in the range 5--10~kpc, i.e. in the region corresponding to values of the distance inferred by \cite{Kuiper_20}, that is \textit{d}~=~(7.6$\pm$0.7)~kpc. Assuming the value of N$_H$ found by \cite{Kuiper_20}, the value of expected extinction is A$_{K_s}$=0.59$\pm$0.01~mag, corresponding to a distance of $d=(7.2\pm0.8)$~kpc.

The estimation of the distance through the hydrogen column density is strongly affected by uncertainties on this parameter. The accuracy in the evaluation of the column density depends on the quality of the data and on the energy range in which the fit is conducted. A small energy range and a low resolution can lead to correlation effects with other parameters of the model. The use of solar abundances as initial values for the parameters can also lead to differences around 5\% in the estimation \citep[see][]{Wilms2000}. Moreover, the accuracy extinction map itself depends on the quality of the data used, whether they are recent data with higher quality, and the analysed region of the sky, which can be affected by strong intrinsic variations in the interstellar medium. Nonetheless, the two estimates of the distance, by means of the PRE analysis of \citet{Kuiper_20} and through the spectral evolution of this work, are compatible.

\section{Conclusions}
\label{sec:conclusion}

We analysed a large sample of the available observations of \source in the X-ray archive with the aim of characterising the spectral emission of the source and its evolution during the outburst. 
The source is well fitted by an absorbed Comptonisation component that on average contributes to 95\% of the whole budget of the unabsorbed emitted flux. No significant spectral changes were found for the source during the whole outburst. The estimation of the distance is in line with the same values reported in literature, and equal to $d=(7.2\pm0.8)$~kpc.

The spectral continuum, especially in the time window at which the flux reaches higher values during the outburst, needs to be fitted with a black-body component peaking at a temperature that is correlated with the \nicer count-rate. This component is characterised by a temperature of about 0.8 keV and appears to be emitted from a region with a radius of (3.3$\pm$1.5)~km that could be compatible with a fraction of the NS surface, or possibly with the boundary layer. No significant variations are observed on the electron temperature of the Comptonising cloud that mainly shows a temperature of about 34 keV. The corona appears to be characterised by an optical depth of about 2.3, which could in part explain the direct black-body component that could arise from a fraction of photons emitted by a small region of the NS surface and that are not significantly scattered in the corona. 
The spectral continuum appears to be characterised by a \ion{Ne}{ix} emission line. This line, however, seems to have a broad profile that is compatible or some times higher than the spectral resolution of \nicer at 1 keV. This could be produced by relativistic effects, if this feature is originated in the innermost parts of the accretion disc (the evidence of which is not detectable from the analysed data), or could be an effect of the spectral resolution of \nicer, being unable to resolve a complex of lines at those energies.

A broad iron emission line has been detected in the spectra of a number of observations. The energy line appears to be correlated to the phase of the outburst. In particular, for lower flux regimes, we observe a broad ($\sim 1$ keV) \ion{Fe}{i} K$\alpha$ line, while in proximity of the peak of the outburst we observed a \ion{Fe}{xxv} K$\alpha$ line. 
Observations of future outbursts of \source with instruments equipped with larger effective area over a wider energy range, as for example the enhanced X-ray Timing and Polarimetry mission \citep[eXTP, ][]{extp,extp_observatory}, could provide important constraints on the possible reflection component in this system, and then also on the ionization state of the matter and on the inclination angle of the system that can not be inferred by the current statistics.

\section*{Acknowledgements}

The authors acknowledge financial contribution from the agreement ASI-INAF n.2017-14-H.0 from INAF mainstream (PI: A. De Rosa), and from the HERMES project financed by the Italian Space Agency (ASI) Agreement n. 2016/13 U.O and from the ASI-INAF Accordo Attuativo HERMES Technologic Pathfinder n. 2018-10-H.1-2020. We also acknowledge support from the European Union Horizon 2020 Research and Innovation Framework Programme under grant agreement HERMES-Scientific Pathfinder n. 821896.
RI and TDS acknowledge the research grant iPeska (PI: Andrea Possenti) funded under the INAF national call Prin-SKA/CTA approved with the Presidential Decree 70/2016. RI acknowledges financial contribution from the agreement ASI-INAF n.2017-14-H.0, from INAF mainstream (PI: T. Belloni).
A. Marino is supported by the H2020 ERC Consolidator Grant "MAGNESIA" under grant agreement No. 817661 (PI: Rea) and National Spanish grant PGC2018-095512-BI00. This work was also partially supported by the program Unidad de Excelencia Maria de Maeztu CEX2020-001058-M, and by the PHAROS COST Action (No. CA16214).

\section*{Data availability}
The data utilized in this article are publicly available in the Heasarc Data Archive at \url{https://heasarc.gsfc.nasa.gov/cgi-bin/W3Browse/w3browse.pl}. The \astro data are publicly available in the ISRO Science Data Archive at \url{https://webapps.issdc.gov.in/astro_archive/archive/Search.jsp}. 





\bibliographystyle{mnras}
\bibliography{biblio}




\appendix

\section{Tables}

\clearpage

\onecolumn{

\begin{longtable}{cccc} 
\caption{Observations considered in this work.}
\footnotesize								
\centering								
\tabularnewline								
\toprule

ObsID	&	Satellite	&	Start Time (UT)	&	Stop Time (UT)	\\	
								
\midrule								
								
90401331002	&	NuSTAR	&	2018-08-13T22:36:09	&	2018-08-14T14:26:09	\\	
1200310101	&	NICER	&	2018-08-14T23:59:42 	&	2018-08-15T14:08:00 	\\	
80301311002	&	NuSTAR	&	2018-08-17T20:01:09	&	2018-08-18T13:21:09	\\	
Rev. 1989	&	INTEGRAL	&	2018-8-17 10:45:21	&	2018-8-19 19:44:37	\\	
1200310102	&	NICER	&	2018-08-18T02:07:53 	&	2018-08-18T03:56:20 	\\	
1200310103	&	NICER	&	2018-08-23T00:58:20 	&	2018-08-23T23:01:36 	\\	
9000002320	&	AstroSAT	&	2018-08-23T01:10:15	&	2018-08-24T00:41:18	\\	
20173	&	Chandra	&	2018-08-23T17:40:05	&	2018-08-23T23:31:42	\\	
1200310104	&	NICER	&	2018-08-24T01:51:10 	&	2018-08-24T14:39:23 	\\	
Rev. 1992	&	INTEGRAL	&	2018-8-25 13:08:59	&	2018-8-27 19:11:30	\\	
1200310105	&	NICER	&	2018-08-27T00:49:37 	&	2018-08-27T22:39:23 	\\	
9000002332	&	AstroSAT	&	2018-08-27T00:05:56	&	2018-08-27T21:59:56	\\	
1200310106	&	NICER	&	2018-08-27T23:55:20 	&	2018-08-28T01:44:44 	\\	
1200310107	&	NICER	&	2018-08-30T15:18:00 	&	2018-08-30T17:35:00 	\\	
Rev. 1994	&	INTEGRAL	&	2018-8-30 20:48:49	&	2018-9-2 3:31:23	\\	
1200310108	&	NICER	&	2018-08-31T22:13:38 	&	2018-08-31T22:55:20 	\\	
1200310109	&	NICER	&	2018-08-31T23:47:15 	&	2018-09-01T17:05:18 	\\	
1200310110	&	NICER	&	2018-09-02T06:40:10 	&	2018-09-02T23:58:36 	\\	
Rev. 1995	&	INTEGRAL	&	2018-9-2 12:39:01	&	2018-9-4 19:22:19	\\	
795750101	&	XMM-Newton	&	2018-09-03T18:47:34.08	&	2018-09-04T03:58:54.86	\\	
1200310111	&	NICER	&	2018-09-03T01:12:30 	&	2018-09-03T07:41:56 	\\	
1200310112	&	NICER	&	2018-09-05T17:59:40 	&	2018-09-05T21:55:00 	\\	
1200310113	&	NICER	&	2018-09-06T00:24:00 	&	2018-09-06T06:49:00 	\\	
1200310114	&	NICER	&	2018-09-07T08:42:40 	&	2018-09-07T09:26:40 	\\	
1200310115	&	NICER	&	2018-09-08T01:42:11 	&	2018-09-08T11:42:51 	\\	
1200310116	&	NICER	&	2018-09-09T03:58:11 	&	2018-09-09T22:59:33 	\\	
Rev. 1998	&	INTEGRAL	&	2018-9-10 12:07:30	&	2018-9-12 18:48:26	\\	
1200310118	&	NICER	&	2018-09-11T00:44:36 	&	2018-09-11T17:58:50 	\\	
1200310119	&	NICER	&	2018-09-11T23:54:45 	&	2018-09-12T21:49:14 	\\	
1200310120	&	NICER	&	2018-09-13T05:15:23 	&	2018-09-13T22:53:29 	\\	
1200310121	&	NICER	&	2018-09-14T07:57:04 	&	2018-09-14T17:04:00 	\\	
1200310122	&	NICER	&	2018-09-15T00:54:21 	&	2018-09-15T19:31:40 	\\	
Rev. 2001	&	INTEGRAL	&	2018-9-18 11:33:28	&	2018-9-20 18:17:35	\\	
1200310124	&	NICER	&	2018-09-19T06:26:49 	&	2018-09-19T22:06:40 	\\	
1200310125	&	NICER	&	2018-09-20T05:35:18 	&	2018-09-20T15:09:00 	\\	
Rev. 2002	&	INTEGRAL	&	2018-9-21 4:47:37	&	2018-9-23 10:06:50	\\	
1200310126	&	NICER	&	2018-09-21T00:07:18 	&	2018-09-21T11:14:40 	\\	
1200310127	&	NICER	&	2018-09-23T07:45:04 	&	2018-09-23T11:11:45 	\\	
Rev. 2003	&	INTEGRAL	&	2018-9-23 19:12:26	&	2018-9-26 1:57:06	\\	
1200310128	&	NICER	&	2018-09-24T02:17:04 	&	2018-09-24T02:41:20 	\\	
1200310129	&	NICER	&	2018-09-25T01:27:04 	&	2018-09-25T15:49:20 	\\	
1200310130	&	NICER	&	2018-09-26T03:55:00 	&	2018-09-26T04:11:20 	\\	
Rev. 2004	&	INTEGRAL	&	2018-9-26 11:02:00	&	2018-9-28 17:47:47	\\	
1200310131	&	NICER	&	2018-09-28T00:30:47 	&	2018-09-28T05:38:00 	\\	
Rev. 2005	&	INTEGRAL	&	2018-9-29 4:14:04	&	2018-10-1 9:38:30	\\	
Rev. 2006	&	INTEGRAL	&	2018-10-1 18:42:44	&	2018-10-4 0:28:03	\\	
1200310132	&	NICER	&	2018-10-04T01:52:19 	&	2018-10-04T22:15:00 	\\	
1200310133	&	NICER	&	2018-10-05T18:10:40 	&	2018-10-05T21:19:40 	\\	
1200310134	&	NICER	&	2018-10-10T20:33:44 	&	2018-10-10T20:48:40 	\\	
1200310135	&	NICER	&	2018-10-11T01:11:02 	&	2018-10-11T01:28:20 	\\	
1200310136	&	NICER	&	2018-10-12T06:32:14 	&	2018-10-12T06:56:18 	\\	
1200310137	&	NICER	&	2018-10-13T13:24:10 	&	2018-10-13T15:15:40 	\\	
1200310138	&	NICER	&	2018-10-16T03:14:20 	&	2018-10-16T05:06:20 	\\	
1200310139	&	NICER	&	2018-10-17T05:32:22 	&	2018-10-17T18:13:20 	\\	
								
\bottomrule								
\end{longtable}

\setlength{\tabcolsep}{6pt} 
\renewcommand{\arraystretch}{1.3} 

\begin{landscape}																							
\footnotesize																					
\begin{longtable}{cccccccccccc} 																							
\caption{Best fit model continuum for each observation of \source. The associated errors are reported at 90\% confidence level.}																							
\tabularnewline																							
\cline{3-12}																							
	&		&	\multicolumn{1}{|c|}{\sc Tbabs}	&	\multicolumn{2}{c|}{\sc Edge}			&	\multicolumn{2}{c|}{\sc Bbodyrad}			&	\multicolumn{4}{c|}{\sc nthComp}							&	\multicolumn{1}{c|}{$\chi^{2}/dof$}	\\
\midrule																							
ObsID	&	MJD	&	N$_{H}$	&	EdgeE	&	MaxTau	&	kT	&	Norm	&	Gamma	&	$kT_{e}$	&	kT$_{seed}$	&	Norm	&		\\
	&		&	 ($\times 10^{22}$)	&	(keV)	&		&	(keV)	&		&		&	(keV)	&	(keV)	&		&		\\
\midrule																							
1200310101	&	58345.0	&	2.09*	&	$0.871*$	&	$2.0\pm0.2$	&	$1.10\pm0.06$	&	$0.8\pm0.2$	&	$1.76\pm0.02$	&	$22^{+4}_{-3}$	&	$0.43\pm0.03$	&	$0.0153^{+0.0010}_{-0.0009}$	&	$2444.3/2147$	\\
1200310102	&	58348.09	&	2.09*	&	$0.871*$	&	$2.5^{+0.6}_{-0.2}$	&	--	&	--	&	$1.77\pm0.02$	&	$22^{+8}_{-4}$	&	$1.03^{+0.06}_{-0.07}$	&	$0.0019^{+0.0003}_{-0.0002}$	&	$1541.3/1515$	\\
1200310103	&	58353.04	&	2.09*	&	$0.871*$	&	$2.0\pm0.1$	&	--	&	--	&	1.9*	&	38.8*	&	$0.58\pm0.01$	&	$0.0150\pm0.0004$	&	$1507.4/1641$	\\
1200310104	&	58354.08	&	2.09*	&	$0.871*$	&	$1.7\pm0.2$	&	--	&	--	&	$1.91\pm0.05$	&	38.8*	&	$0.63\pm0.02$	&	$0.0139^{+0.0007}_{-0.0006}$	&	$1346.2/1563$	\\
1200310105	&	58357.03	&	2.09*	&	$0.871*$	&	$2.2\pm0.3$	&	$1.02^{+0.08}_{-0.09}$	&	$4.4^{+2.1}_{-1.6}$	&	$1.65^{+0.07}_{-0.07}$	&	38.8*	&	$0.44^{+0.06}_{-0.07}$	&	$0.016\pm0.002$	&	$760.8/718$	\\
1200310106	&	58358.0	&	2.09*	&	$0.871*$	&	$2.5^{+0.3}_{-0.7}$	&	$1.15^{+0.09}_{-0.07}$	&	$3\pm1$	&	$1.71^{+0.07}_{-0.06}$	&	$40^{+57}_{-13}$	&	$0.42^{+0.10}_{-0.13}$	&	$0.019^{+0.009}_{-0.004}$	&	$538.0/557$	\\
1200310107	&	58360.64	&	2.09*	&	--	&	--	&	--	&	--	&	$1.82\pm0.05$	&	$31^{+55}_{-10}$	&	$0.54\pm0.02$	&	$0.012\pm0.002$	&	$657.4/623$	\\
1200310108	&	58361.93	&	2.09*	&	$0.871*$	&	$2.1\pm0.3$	&	--	&	--	&	$1.84^{+0.07}_{-0.06}$	&	$32^{+86}_{-11}$	&	$0.55\pm0.04$	&	$0.022\pm0.002$	&	$587.1/540$	\\
1200310109	&	58361.99	&	2.09*	&	$0.871*$	&	$1.8\pm0.2$	&	--	&	--	&	$1.90\pm0.04$	&	$\geq 42$	&	$0.60\pm0.02$	&	$0.0207^{+0.0009}_{-0.0008}$	&	$798.7/769$	\\
1200310110	&	58363.28	&	2.09*	&	$0.871*$	&	$1.9\pm0.2$	&	$0.58\pm^{+0.04}_{-0.05}$	&	$58^{+10}_{-8}$	&	$2.3^{+0.3}_{-0.2}$	&	38.8*	&	$1.1\pm0.2$	&	$0.0061^{+0.0028}_{-0.0015}$	&	$865.2/745$	\\
1200310111	&	58364.05	&	2.09*	&	$0.871*$	&	$1.6\pm0.2$	&	--	&	--	&	$1.79\pm0.02$	&	38.8*	&	$0.57\pm0.02$	&	$0.0164\pm0.0008$	&	$1190.3/840$	\\
1200310112	&	58366.75	&	2.09*	&	$0.871*$	&	$1.4\pm0.6$	&	$0.61^{+0.09}_{-0.10}$	&	$36^{+21}_{-8}$	&	1.9*	&	38.8*	&	$1.3^{+0.5}_{-0.4}$	&	$\leq 0.0023$	&	$533.0/489$	\\
1200310113	&	58367.02	&	2.09*	&	$0.871*$	&	$2.4\pm0.4$	&	--	&	--	&	$1.78^{+0.06}_{-0.05}$	&	38.8*	&	$0.45\pm0.05$	&	$0.017\pm0.002$	&	$561.6/535$	\\
1200310114	&	58368.36	&	2.09*	&	$0.871*$	&	$1.4\pm0.3$	&	--	&	--	&	1.9*	&	38.8*	&	$0.60\pm0.03$	&	$0.0132^{+0.0010}_{-0.0009}$	&	$492.6/496$	\\
1200310115	&	58369.07	&	2.09*	&	$0.871*$	&	$1.9\pm0.2$	&	--	&	--	&	$1.87^{+0.06}_{-0.05}$	&	38.8*	&	$0.57\pm0.03$	&	$0.0148^{+0.0010}_{-0.0009}$	&	$683.9/643$	\\
1200310116	&	58370.17	&	2.09*	&	$0.871*$	&	$2.0\pm0.2$	&	--	&	--	&	$1.93^{+0.06}_{-0.05}$	&	$\geq 41$	&	$0.62\pm0.03$	&	$0.0168^{+0.0013}_{-0.0009}$	&	$735.3/718$	\\
1200310118	&	58372.03	&	2.09*	&	$0.871*$	&	$1.3\pm0.3$	&	$0.73^{+0.08}_{-0.09}$	&	$45^{+8}_{-40}$	&	$1.9\pm0.2$	&	38.8*	&	$1.8^{+0.8}_{-0.5}$	&	$0.005^{+0.014}_{-0.002}$	&	$530.0/499$	\\
1200310119	&	58373.0	&	2.09*	&	$0.871*$	&	$2.0\pm0.3$	&	$0.65^{+0.06}_{-0.10}$	&	$45^{+8}_{-40}$	&	$1.8^{+0.5}_{-0.4}$	&	$\geq 27$	&	$1.2^{+0.4}_{-0.7}$	&	$0.005^{+0.014}_{-0.002}$	&	$719.1/648$	\\
1200310120	&	58374.22	&	2.09*	&	$0.871*$	&	$2.2\pm0.3$	&	--	&	--	&	$1.84\pm0.06$	&	38.8*	&	$0.62\pm0.03$	&	$0.021\pm0.001$	&	$641.6/642$	\\
1200310121	&	58375.33	&	2.09*	&	--	&	--	&	--	&	--	&	$2.19\pm0.02$	&	38.8*	&	$0.87\pm0.04$	&	$0.0130\pm0.005$	&	$561.0/426$	\\
1200310122	&	58376.04	&	2.09*	&	$0.871*$	&	$2.4\pm0.3$	&	--	&	--	&	$1.81^{+0.08}_{-0.07}$	&	38.8*	&	$0.60\pm0.04$	&	$0.025\pm0.002$	&	$601.8/548$	\\
1200310124	&	58380.27	&	2.09*	&	$0.871*$	&	$2.2\pm0.3$	&	--	&	--	&	$1.90^{+0.09}_{-0.07}$	&	$58^{+289}_{-23}$	&	$0.64\pm0.04$	&	$0.028\pm0.002$	&	$511.7/554$	\\
1200310125	&	58381.23	&	2.09*	&	$0.871*$	&	$3.1\pm0.5$	&	--	&	--	&	$1.77^{+0.09}_{-0.07}$	&	$33^{+20}_{-8}$	&	$0.53\pm0.6$	&	$0.033^{+0.005}_{-0.004}$	&	$401.3/420$	\\
1200310126	&	58382.01	&	2.09*	&	$0.871*$	&	$2.2\pm0.2$	&	--	&	--	&	$1.89^{+0.07}_{-0.06}$	&	$40^{+54}_{-13}$	&	$0.62\pm0.03$	&	$0.029\pm0.002$	&	$605.5/621$	\\
1200310127	&	58384.32	&	2.09*	&	$0.871*$	&	$2.1\pm0.3$	&	--	&	--	&	$1.91^{+0.09}_{-0.08}$	&	$33^{+33}_{-10}$	&	$0.63\pm0.04$	&	$0.025\pm0.002$	&	$555.6/561$	\\
1200310128	&	58385.1	&	2.09*	&	$0.871*$	&	$2.0\pm0.3$	&	--	&	--	&	$1.90^{+0.08}_{-0.07}$	&	38.8*	&	$0.61\pm0.04$	&	$0.023\pm0.002$	&	$525.1/523$	\\
1200310129	&	58386.06	&	2.09*	&	$0.871*$	&	$2.1\pm0.3$	&	--	&	--	&	$1.90\pm0.06$	&	38.8*	&	$0.61^{+0.03}_{-0.04}$	&	$0.0215^{+0.002}_{-0.001}$	&	$688.0/630$	\\
1200310130	&	58387.16	&	2.09*	&	$0.871*$	&	$2.0\pm0.4$	&	$0.68^{+0.07}_{-0.08}$	&	$49^{+14}_{-10}$	&	$2.0\pm0.1$	&	38.8*	&	$1.4\pm0.4$	&	$0.004^{+0.003}_{-0.001}$	&	$468.7/465$	\\
1200310131	&	58389.02	&	2.09*	&	$0.871*$	&	$2.0\pm0.3$	&	--	&	--	&	$1.91\pm0.06$	&	38.8*	&	$0.57\pm0.04$	&	$0.019^{+0.002}_{-0.001}$	&	$619.8/613$	\\
1200310132	&	58395.08	&	2.09*	&	$0.871*$	&	$2.1\pm0.4$	&	--	&	--	&	$1.87^{+0.21}_{-0.07}$	&	$\geq 45$	&	$0.45\pm0.05$	&	$0.0168^{+0.003}_{-0.002}$	&	$453.4/453$	\\
1200310133	&	58396.76	&	2.09*	&	--	&	--	&	--	&	--	&	$2.0^{+0.3}_{-0.2}$	&	38.8*	&	$0.62\pm0.07$	&	$0.0075^{+0.0007}_{-0.0006}$	&	$134.4/120$	\\
	
\bottomrule													\end{longtable}						
\label{tab:continuum}
\end{landscape}																
\footnotesize											
\centering											
\begin{longtable}{cccccc} 											
\caption{Best fit parameters for the emission lines detected in the \source spectrum for each observation. The line energy for the \ion{Ne}{ix} ion was fixed to the rest-frame energy, while the sigma was fixed to 0.085 keV for each ion transition (i.e. the spectral resolution of \nicer/XTI at 1 keV). All the associated errors are reported at 68\% confidence level.}											
\tabularnewline											
\cline{3-6}											
	&		&	\multicolumn{1}{|c|}{\ion{Ne}{ix}}	&	\multicolumn{3}{|c|}{\ion{Fe}{xxv}}					\\
\midrule											
ObsID	&	MJD	&	Norm	&	LineE 	&	Sigma	&	Norm	\\
	&		&		&	(keV)	&	(keV)	&		\\
\midrule											
											
1200310101	&	58345.0	&	$0.015\pm0.002$	&	--	&	--	&	--	\\
1200310102	&	58348.09	&	$0.008^{+0.003}_{-0.002}$	&	--	&	--	&	--	\\
1200310103	&	58353.04	&	$0.021\pm0.002$	&	$6.6^{+0.3}_{-0.4}$	&	$0.5\pm 0.3$	&	$0.00015^{0.00006}_{-0.00005}$	\\
1200310104	&	58354.08	&	$0.019^{+0.003}_{-0.002}$	&	--	&	--	&	--	\\
1200310105	&	58357.03	&	$0.0108^{+0.0016}_{-0.0014}$	&	$\leq 6.3$	&	$1.5\pm0.2$	&	$0.0026^{+0.0007}_{-0.0008}$	\\
1200310106	&	58358.0	&	$0.024^{+0.007}_{-0.009}$	&	$6.59\pm0.08$	&	$0.1*$	&	$0.00018^{+0.00002}_{-0.00005}$	\\
1200310107	&	58360.64	&	$0.034^{+0.007}_{-0.006}$	&	--	&	--	&	--	\\
1200310108	&	58361.93	&	$0.028^{+0.007}_{-0.006}$	&	--	&	--	&	--	\\
1200310109	&	58361.99	&	$0.016\pm0.002$	&	--	&	--	&	--	\\
1200310110	&	58363.28	&	$0.016\pm0.002$	&	--	&	--	&	--	\\
1200310111	&	58364.05	&	$0.016^{+0.003}_{-0.002}$	&	$6.7\pm0.2$	&	$0.8^{+0.3}_{-0.2}$	&	$0.00019^{+0.00005}_{-0.00004}$	\\
1200310112	&	58366.75	&	$0.0068^{+0.0019}_{-0.0015}$	&	--	&	--	&	--	\\
1200310113	&	58367.02	&	$0.017^{+0.005}_{-0.004}$	&	--	&	--	&	--	\\
1200310114	&	58368.36	&	$0.016^{+0.004}_{-0.003}$	&	$6.53^{+0.07}_{-0.06}$	&	$0.02^{+0.10}_{-0.02}$	&	$0.00009\pm0.00004$	\\
1200310115	&	58369.07	&	$0.017\pm0.003$	&	--	&	--	&	--	\\
1200310116	&	58370.17	&	$0.019\pm0.003$	&	--	&	--	&	--	\\
1200310118	&	58372.03	&	$0.016^{+0.005}_{-0.004}$	&	--	&	--	&	--	\\
1200310119	&	58373.0	&	$0.019^{+0.004}_{-0.003}$	&	--	&	--	&	--	\\
1200310120	&	58374.22	&	$0.030^{+0.006}_{-0.005}$	&	--	&	--	&	--	\\
1200310121	&	58375.33	&	$0.048^{+0.020}_{-0.014}$	&	--	&	--	&	--	\\
1200310122	&	58376.04	&	$0.041^{+0.011}_{-0.009}$	&	--	&	--	&	--	\\
1200310124	&	58380.27	&	$0.031^{+0.009}_{-0.007}$	&	--	&	--	&	--	\\
1200310125	&	58381.23	&	$0.08^{+0.04}_{-0.03}$	&	--	&	--	&	--	\\
1200310126	&	58382.01	&	$0.033^{+0.007}_{-0.006}$	&	--	&	--	&	--	\\
1200310127	&	58384.32	&	$0.019^{+0.005}_{-0.004}$	&	--	&	--	&	--	\\
1200310128	&	58385.1	&	$0.024^{+0.007}_{-0.005}$	&	--	&	--	&	--	\\
1200310129	&	58386.06	&	$0.019^{+0.004}_{-0.003}$	&	--	&	--	&	--	\\
1200310130	&	58387.16	&	$0.017^{+0.006}_{-0.005}$	&	--	&	--	&	--	\\
1200310131	&	58389.02	&	$0.018^{+0.004}_{-0.003}$	&	--	&	--	&	--	\\
1200310132	&	58395.08	&	$0.013^{+0.004}_{-0.003}$	&	$6.63^{+0.06}_{-0.05}$	&	$0.04331^{+0.09}_{-0.04}$	&	$0.00013^{+0.00005}_{-0.00004}$	\\
1200310133	&	58396.76	&	--	&	--	&	--	&	--	\\
											
\bottomrule											
\end{longtable}											
\label{tab:lines}											
												
}

\bsp	
\label{lastpage}
\end{document}